\documentclass[12pt]{iopart}

\makeatletter
\@namedef{ver@amsmath.sty}{}
\makeatother
\usepackage{amstext}
\usepackage[version=3]{mhchem}
\usepackage[utf8]{inputenc}
\usepackage{amsmath}
\usepackage{setstack}
\usepackage{epstopdf} 
\usepackage{amssymb}
\usepackage{color}
\usepackage[dvips]{graphicx}
\usepackage{subfigure}
\usepackage{color}

\newcommand{\ket}[2][]{{|#2\rangle_{#1}}}
\newcommand{\bra}[2][]{{}_{#1}\langle #2|}
\def\ben{\begin{eqnarray}}
\def\een{\end{eqnarray}}
\def\bei{\begin{itemize}}
\def\eei{\end{itemize}}

\def\<{\langle}
\def\>{\rangle}

\begin{document}
\def\be{\begin{equation}}
\def\ee{\end{equation}}
\newtheorem{thm}{Theorem}
\newtheorem{prop}[thm]{Proposition}
\newtheorem{cor}[thm]{Corollary}
\newtheorem{lem}[thm]{Lemma}
\newtheorem{fact}[thm]{Fact}
\newtheorem{remark}[thm]{Remark}

\title{Information transfer during the universal gravitational  decoherence}

\author{J.~K.~Korbicz $^{1,2}$, J.~Tuziemski$^{1,2}$}
\address{$^1$Faculty of Applied Physics and Mathematics, Gda\'nsk University of Technology, 80-233 Gda\'nsk, Poland}
\address{$^2$National Quantum Information Centre in Gda\'nsk, 81-824 Sopot, Poland}
\ead{jkorbicz@mif.pg.gda.pl}

\begin{abstract}
	Recently Pikovski et al. have proposed in [ Pikovski I et al. 2015  \emph{Nature Phys.} {\bf 11}, 668] an intriguing universal decoherence mechanism, suggesting  that gravitation may play
	a conceptually important role in the quantum-to-classical transition, albeit vanishingly small in everyday situations.  
	Here we analyze information transfer 
	induced by this mechanism. We show that generically on short time-scales, gravitational decoherence leads to
	a redundant information encoding, which results in a form of  objectivization of the center-of-mass position in the gravitational field. 
      We derive the relevant time-scales of this process, given in terms of energy dispersion and quantum Fisher information.  
	As an example we study thermal coherent states and show certain robustness of the effect with the temperature. 
      Finally, we draw an analogy between our objectivization mechanism and the  fundamental problem of point individuation in General Relativity as emphasized by the Einstein's Hole argument.

\end{abstract}

\maketitle

\section{Introduction}

Emergence of the classical world from quantum has been a long standing problem. Theory of decoherence \cite{decoh} is one of the attempts to resolve it.   
In the recent paper \cite{pikovski}, Pikovski et al. proposed a universal decoherence mechanism due to the gravitational time dilation \cite{gravitation}.
Unlike some of the earlier proposals, e.g.  due to Di\'osi \cite{diosi}  or  Penrose \cite{penrose}, the mechanism of Pikovski et al. does not require an immediate departure from unitarity. 
It has generated a lively debate as to weather the effect is real \cite{bonder1, bonder2, diosi2, qa} or is it really gravity-related \cite{diosi2, pang}.
In this work we will not enter this discussion but rather assume the mechanism of \cite{pikovski} to be true and examine the gravity-mediated
information transfer it entails.

Let us recall the idea behind the mechanism of Pikovski et al. \cite{pikovski}. In gravitational field time flows differently depending on the position in the field,
which leads to a position-dependent gravitational redshift of frequencies. This effect has been one of the classical tests of general relativity, starting with the experiment of Pound and Rebka \cite{pr},
and has been recently confirmed over an astonishingly small height difference of  $33$cm in the Earth gravitational field \cite{wineland}.
Thus, systems which have some natural frequencies associated with their dynamics, like harmonic oscillators, will effectively couple to the position in the gravitational field.
If we now promote this reasoning to the quantum domain, then the position-frequency correlation will in general lead to a position decoherence if (some of) the oscillatory degrees of freedom are
left unobserved. This, in turn,  will lead to a loss of visibility, if one performs an interferometric experiment, and this loss can be directly related to the proper time difference at different heights.
Since the gravitational coupling is weak, the effect requires a macroscopic
amount of oscillators $N\sim 10^{23}$ to obtain reasonable decoherence rates (milliseconds for micrometric height differences) \cite{pikovski}. A surprising aspect of the mechanism of Pikovski et al.
is that there is no position information storage in the local oscillator degrees of freedom (which we will simply call the environment) if the latter are initially in a thermal equilibrium. 
Thus, decoherence happens without a localized which-path information in the sense that a reduced state of any portion of the environment is unchanged by the evolution,
and for short times is driven by the internal energy dispersion \cite{pikovski, adler}.

However, as pointed out by Zurek \cite{ZurekNature}, decoherence alone is generally not enough for the emergence of such an important aspect of the classical world as its objective character.
Briefly speaking, one has to show that during the decoherence process there is a redundant transfer of information into the environment and this information is accessible to multiple observers.
Here, we perform such studies in the universal gravitational decoherence model and show that many identical copies of the position information can be deposited in the environment if the latter is in a non-stationary state.
Our main tool are the so called Spectrum Broadcast Structures (SBS) \cite{prl, pra}---an approach evolved from
the quantum Darwinism idea \cite{ZurekNature, pawel} and based on direct studies of, what we call, an extended state. This is a quantum state of the system plus a part of its environment
and its analysis goes beyond the standard methods of the decoherence theory (e.g. the Master equation techniques), which deal with the reduced state of the system only. 
Using the simplified model of \cite{pikovski}, we solve for the extended state and show that on short time-scales
there are general regimes where it approaches the SBS form. This in turn, implies that the universal gravitational decoherence  leads to a redundant position information storage in the environment and hence its 
objectification \cite{ZurekNature,pra}. 
The efficiency of this storage is governed by the Quantum Fisher Information (QFI) (see e.g. \cite{geza}), analogously
to decoherence being dependent on the energy dispersion \cite{pikovski}. 
Going beyond the general short-time analysis (cf. \cite{adler,jp}), we study in detail an example of the internal degrees of freedom prepared in thermal coherent states and the information transfer efficiency  as a function of the temperature.
We study the formation of the SBS and as a by-product derive a form of an information gain -- disturbance  uncertainty relation.

An information storage in the internal degrees of freedom for interference experiments in gravitational field has been already studied, e.g. in \cite{natcom}.  However, 
the internal degrees of freedom were treated there in a standard for open quantum systems way -- were traced out and their effect on the visibility studied. 
This approach cannot show how exactly information about the system is proliferated in the environment, e.g. if there are many records of the same information being formed
and information becomes objective. Here we 
present a more refined analysis, dividing the internal degrees of freedom into observed and unobserved parts, which allows to study such questions.

Let us also mention  that there exist some general results within the quantum Darwinism  \cite{Zwolak} and SBS  \cite{pomiary} approaches concerning, so called, pure decoherence Hamiltonians considered also here,
and suggesting that the whole class leads to the emergence of objectivity. However, in the first work, based on quantum Chernoff information,  the result is strictly speaking derived only 
for two-dimensional central systems and is in part based on somewhat heuristic typicality arguments, concerning quantum Chernoff information. 
Moreover \cite{Zwolak} provides no timescales of the objectification process.
The second mentioned result \cite{pomiary} is in turn based solely on statistical arguments, the derived time-scales are very general in nature, and is again bounded to finite dimensions only.  
Thus, to convincingly prove the appearance of objectivity during the universal gravitational decoherence, one has to study the model in more detail from the quantum-information perspective. 
This is what we do here. We end our work with an intriguing and rather speculative analogy between the gravity-induced information transfer and the, so called, 
point individuation in General Relativity (see e.g. \cite{stachel1} for a modern review).

\section{Extended state dynamics under the universal gravitational decoherence}
Let us begin with our main tool - Spectrum Broadcast Structures (SBS). We briefly introduce them below, referring the reader to e.g. \cite{prl,pra} for more information. 
SBS is defined as a multipartite quantum state of the system plus a portion of its environment of the following form:
\be\label{SBS0}
\varrho_{SBS}= \sum_ip_i\ket i\bra i \otimes \varrho^{(1)}_i\otimes\cdots\otimes\varrho^{(M)}_i, 
\ee
with 
\be\label{perp}
\varrho^{(k)}_i\perp\varrho^{(k)}_{i'\ne i}.
\ee
Here $\ket i$ are the so called pointer states into which the system (the center of mass here) has decohered, $p_i$ are the initial pointer probabilities,
and $\varrho^{(k)}_i$ are states of the fragments of the internal degrees of freedom. The condition (\ref{perp}), meaning that the states $\varrho^{(k)}_i$ 
have orthogonal supports for different pointer index $i$, ensures that they are one shot perfectly distinguishable. This, together with (\ref{SBS0}),  implies that the same state of the decohered 
CM, labeled by the index $i$, can be read off from many portions of the environment without any disturbance (on average) to the extended state.
This, in turn leads, to a form of an emergent objectivity \cite{ZurekNature,pra} of the state of the system: All the observers will see the same state of the system without disturbing neither the state  nor each other.
Conversely, as proved under certain assumptions in \cite{pra},  it  can be shown that demanding objectivity, in the sense of information redundancy and 
non-disturbance in its extraction, leads to SBS. In this sense SBS are responsible for some form of classicality - the emergent objectivity. 

Our main goal is to show that during the universal gravitational decoherence there are regimes such that the extended state of the center-of-mass (CM) of a decohering particle 
plus a fraction of its internal 
degrees of freedom (serving as an environment) approaches the SBS form (\ref{SBS0})-(\ref{perp}).
Following \cite{pikovski}, we consider a compound system with a large number of internal degrees of freedom, effectively described by uncoupled harmonic oscillators
(e.g. a large molecule), placed in a gravitational filed. We separate the motion into the center-of-mass motion and the internal oscillations.
With such a division, the CM is treated as a (virtual) central system and the internal oscillators as its environment. In this work, contrary to the previous studies \cite{pikovski, diosi2,adler},
we will assume that only a part of the internal oscillators can be traced out, while the other is left for an observation. 
This in fact represents a very common situation of an indirect observation \cite{ZurekNature} and forces us to study the extended state of the 
central system plus the observed part of its environment \cite{prl,pra}.
Assuming applicability of the energy-mass relation to the Klein-Gordon equation in a weak gravitational field and non-relativistic velocities, 
the corresponding approximate Hamiltonian  was derived in \cite{lamerzahl, pikovski}:
\be
H_{tot}=H_{cm}(X,P)+H_0\left(1+\frac{\Phi(X)}{c^2}-\frac{P^2}{2m^2c^2}\right),
\ee
where $H_{cm}$ is a center-of-mass Hamiltonian depending on the canonical variables $X,P$, $H_0$ is the internal (oscillatory) degrees of freedom Hamiltonian, and 
$\Phi(X)$ is the Newtonian gravitational potential. The crucial step is now an assumption that this is a valid Hamiltonian in the 1st quantization too, symbolically $H_{tot}\to\hat H_{tot}$.
Some further simplifications can be made \cite{pikovski}: i) neglecting $H_{cm}$ as compared to the rest; ii) neglecting the special-relativistic kinematical term  $P^2/2M^2c^2$;
iii) assuming a homogeneous  gravitational field (e.g. from the Earth) $\Phi(X)=gX$, where $g$ is the gravitational acceleration.
This finally leads to the Hamiltonian: 
\begin{eqnarray}
\hat H_{tot}\approx \hat H_{0}+\hat H_{int}= \hat H_{0}+\Phi(\hat X)\otimes \frac{\hat H_0}{c^2}.
\end{eqnarray}
Let us stress again that although motivated by the field-theoretical arguments \cite{pikovski} this is a quantum mechanical Hamiltonian and we work in the 1st quatization.
As we already mentioned, $H_0$ describes just a collection of $N$ independent oscillators (one can think of degrees of freedom of a large molecule \cite{pikovski}) so that (we neglect the zero-point energy): 
\be
\hat H_0\equiv\sum_{i=k}^{N}\hat H_{0k}=\sum_{i=k}^{N}\hbar\omega_k \hat n_k, \, \hat n_k=\hat a_k^\dagger \hat a_k.
\ee 
This evolution can be easily solved, recalling that $\Phi(\hat X)=g\hat X$ and writing with the full formality:

\be
\hat H_{tot}={\bf 1}\otimes H_0 + \hat X\otimes \frac{g\hat H_0}{c^2}=\int dX \ket X\bra X \otimes \sum_k \hbar\left(1+\frac{gX}{c^2}\right)\omega_k\hat n_k,
\ee
where we used the expansions ${\bf 1}=\int dX \ket X\bra X$, $\hat X = \int dX X \ket X\bra X$. Calculation of the exponential series of the above Hamiltonian is easy due to the 
orthogonality of the position eigenstates. This finally gives the full system evolution:
\be\label{U}
\hat U(t)\equiv \textrm e^{-i t\hat H_{tot}/\hbar}=\int dX\ket X\bra X\otimes \left[\bigotimes_{k=1}^N\textrm e^{-it\omega_k(X)\hat n_k}\right],
\ee

where 
\be
\omega_k(X)\equiv \left(1+\frac{gX}{c^2}\right)\omega_k
\ee
are the red-shifted frequencies. We apply (\ref{U}) to an initial state, assumed to be
$\varrho_{tot}(0)=\varrho_0\otimes \bigotimes_i\varrho_{0i}$, and divide the oscillators into two fractions: the observed one 
of a size $N_o$ and the unobserved, containing $N_\perp$ oscillators, $N_o+N_\perp=N$. The exact sizes are not important for our considerations, apart
from $N_o,N_\perp$  scaling with the total number $N$ (this defines what we call a macrofraction), 
as we will be tacitly assuming a thermodynamic-type of a limit $N\to\infty$ \cite{prl}.
Tracing the unobserved part of the environment, gives what we call the extended state and what will be our main object of the study:
\begin{eqnarray}
\label{mama}
&&\varrho_{ext}(t)\equiv\tr_{uno} \varrho_{tot}=\\ \nonumber
&&=\int\int dX dX' \varrho_0(X,X') \Gamma_t(\Delta X)\ket X\bra {X'}\otimes \left[ \bigotimes_{k\in N_{o}} \hat U^{(k)}_t(X)\varrho_{0k}\hat U^{(k)}_t(X')^\dagger \right],
\end{eqnarray}
where $\varrho_0(X,X')\equiv \bra{X'}\varrho_0\ket X$, 
\be\label{Ui}
\hat U^{(k)}_t(X)\equiv\text{exp}[-i\omega_k(X)t\hat n_k],
\ee 
and
\begin{eqnarray}
&&\Gamma_t(\Delta X) \equiv\prod_{k=1}^{N_\perp}\tr\left[\hat U^{(k)}_t(X')^\dagger\hat U^{(k)}_t(X)\varrho_{0k}\right]=\prod_{k=1}^{N_\perp}\tr\left[\text e^{i\frac{g\Delta X \omega_k t}{c^2}\hat n_k}\varrho_{0k}\right]\label{G}
\end{eqnarray}
is what can be called a partial decoherence factor \cite{ZurekNature}. Unlike in standard approaches to open quantum systems, which is also the case of \cite{pikovski}, here it arises due to tracing out only a fraction of oscillators (their number denoted $N_\perp$)
and not the whole environment. It controls a trace-norm decay of off-diagonal, in the position basis of the CM, operator terms in $\varrho_{ext}(t)$ \cite{prl,piotrek}. We note that it depends only on the height separation $\Delta X\equiv X'-X$.

\section{General short-time approximation: The role of Quantum Fisher Information and energy dispersion}
Let us first perform a short-time analysis \cite{pikovski,adler} for times such that all the phase differences in (\ref{G}) are small:
\be\label{st}
\Delta\phi_k\equiv \frac{g|\Delta X|}{c^2}\omega_k t\ll1
\ee
or equivalently for times much smaller than the time-dilation induced change of the oscillation periods. This is the basic approximation we will use in the following two Sections.
Decoherence factor in this approximation has been found in \cite{pikovski, adler} and reads:
\be\label{G_st}
\left|\Gamma_t(\Delta X)\right|\approx\text{exp}\left[-\left(\frac{g\Delta X }{\sqrt 2 \hbar c^2}\right)^2\left(\sum_{i=1}^{N_\perp}\Delta  H_{0i}^2\right)t^2\right],
\ee
where $\Delta H_{0i}^2\equiv \tr(\varrho_{0i}\hat H_{0i}^2) - [\tr(\varrho_{0i}\hat H_{0i})]^2$ is the variance of the $i$-th oscillator energy, calculated in the initial state $\varrho_{0i}$.
Let us assume that the amount of the unobserved oscillators is very large so that we may use the Law of Large Numbers to write the sum above in a more compact form as an ensemble average over the
unobserved oscillators. This is somewhat similar e.g. to the usual introduction of a spectral density to describe the environment \cite{decoh}.  Let the 
unobserved frequencies
$\omega$ be distributed with some probability $p_\perp(\omega)$ such that the average 
$\langle\langle \Delta H_{0}^2\rangle\rangle\equiv \int d\omega p_\perp(\omega) \Delta H_{0\omega}^2$ exists (note that in general 
the initial states are also $\omega$-dependent). Then:
\be\label{Gst}
\left|\Gamma_t(\Delta X)\right|\approx\text{exp}\left[-\frac{N_\perp g^2\Delta X^2 \langle\langle\Delta H_{0}^2\rangle\rangle}{2 \hbar^2 c^4}t^2\right],
\ee
giving the decoherence time in a form resembling the energy-time uncertainty relation:
\be\label{tdec}
\tau_{dec}\sqrt{N_\perp\langle\langle\Delta H_{0}^2\rangle\rangle}\equiv \frac{\sqrt 2\hbar c^2}{g|\Delta X|}.
\ee
This decoherence time will be within the validity of the used approximation (\ref{st}) if for all the relevant frequencies $\omega$ the cumulative energy variance is big enough:
\be
N_\perp\langle\langle\Delta H_{0}^2\rangle\rangle\gg \left(2\hbar\omega\right)^2.\label{var}
\ee

Let us now study the distinguishability conditions (\ref{perp}). The post-interaction states of the internal oscillators (we recall that some of them are left for observation and thus cannot be traced out) read:
\be
\varrho_t^{(i)}(X)\equiv U^{(i)}_t(X)\varrho_{0i}U^{(i)}_t(X)^\dagger.\label{rhot} 
\ee
We are interested in the information they carry about the position $X$, as measured by their distinguishability
Since $U^{(i)}_t(X)$ is generated by the red-shifted local Hamiltonian $\hat H_{0i}(X)$, it is obvious that 
any initial state $\varrho_{0i}$ preserved by $\hat H_{0i}(X)$, $[\varrho_{0i},\hat H_{0i}(X)]=0$, will encode no information at all
and this is the situation of \cite{pikovski}. However, here we assume a generic, mixed $\varrho_{0i}$. 
As a measure of state distinguishability we choose the state fidelity (also called generalized overlap or Bhattacharyya coefficient) 
$B(\varrho,\sigma)\equiv\tr\sqrt{\sqrt\varrho\sigma\sqrt\varrho}$ \cite{fuchs, prl}.
There are of course other measures, for example quantum Chernoff information, providing a tighter
than the fidelity upper bound on the probability of discrimination error (see e.g. \cite{Chernoff}). However, as we are ultimately
interested in a perfect distinguishability (cf. (\ref{perp})), the weaker character of the bound is not so important as $B\to 0$
implies that both quantum Chernoff information and the probability of error vanish (in fact $B$ provides also a lower bound
for the probability of error \cite{fuchs}). As $B$ is easier to calculate (no optimization involved, cf. \cite{Chernoff}), we choose to work with it.
Let us consider the fidelity:
\be
B^i_t(\Delta X)\equiv B[\varrho_t^{(i)}(X),\varrho_t^{(i)}(X')], 
\ee
which is a function of  the separation $\Delta X$ only, since  the states $\varrho_t^{(i)}(X)$ differ by an unitary rotation depending on $\Delta \phi_i$ (\ref{st}):
\be
B[\varrho_t^{(i)}(X),\varrho_t^{(i)}(X')]=\tr\sqrt{\sqrt{\rho_{0i}} \text e^{i\Delta \phi_i\hat n_i}\rho_{0i} \text e^{-i\Delta\phi_i\hat n_i} \sqrt{\rho_{0i}}. }
\ee
For short times (\ref{st}), we  develop $B$ in a series in $\Delta X$,  using first the definition of 
the  Bures distance $d_B$  \cite{Zyczkowski_ksiazka} $B(\Delta X)=1-1/2[d_B(\Delta X)]^2$ and then the fact that an infinitesimal 
Bures distance is given by  $1/4$ of the quantum Fisher information (QFI) $F(\varrho_{0i};\hat H_{0i})$ \cite{qfi}. We get the following short-time approximation:
\be\label{b_st}
B^i_t(\Delta X)\approx 1-\frac{t^2}{8}\left(\frac{g\Delta X}{\hbar c^2}\right)^2 F(\varrho_{0i};\hat H_{0i})\approx\text{exp}\left[-\frac{t^2}{8}\left(\frac{g\Delta X}{\hbar c^2}\right)^2 F(\varrho_{0i};\hat H_{0i})\right].
\ee 
By the famous quantum Cramer-Rao bound, $F(\varrho_{0i};\hat H_{0i})$ sets the lower bound on the precision of the experimental estimation of $gt\Delta X/(\hbar c^2)$.
It thus has a very clear operational meaning and a fundamental importance e.g. for the recently popular field of quantum metrology \cite{metrology}. $F(\varrho_{0i};\hat H_{0i})$ can 
be given in terms of the eigenvalues and eigenvectors of the initial state  $\varrho_{0i}=\sum_n \lambda_n \ket{\lambda_n}\bra{\lambda_n}$ 
as  \cite{geza}:
\be\label{qfi}
F(\varrho_{0i};\hat H_{0i})=2\sum_{m,n}\frac{(\lambda_n-\lambda_m)^2}{\lambda_n+\lambda_m} |\langle \lambda_n|\hat H_{0i}| \lambda_m\rangle|^2.
\ee
We see that if  the initial state of the environment $\varrho_{0i}$ is diagonal in the basis of $\hat H_{0i}$, then obviously $F(\varrho_{0i};\hat H_{0i})=0$ and this degree of freedom 
is unable to encode position information about the central system as mentioned earlier. This is an example of a general phenomenon -- If the environment is prepared in a state stationary
w.r.t. the system-environment coupling,  the observers will not be able to decode any
 information about the system using local measurements and
 thus such state will not get objective in the sense discussed here.

Let us now consider more general, and in many cases more realistic, situation where we allow for a grouping of the states (\ref{rhot}) in order to concentrate information they carry. 
This can be viewed as a sort of a  coarse-graining of the internal degrees of freedom \cite{prl} and represent a general situation where each observer has access to a subsystem of the environment rather than a single degree of freedom.
We thus divide the observed part of the internal oscillators, which we stress is a different part of the environment from the one giving rise to decoherence (\ref{G}), into a number of 
smaller fractions, called macrofractions, and denoted $mac_1, \dots, mac_{\mathcal M}$. For simplicity we will assume them to be all of an equal size $N_{mac}$, scaling with the
total number of the observed degrees of freedom $N_{mac}\equiv\mu N_{o}$, $0<\mu<1$. More importantly, we assume $N_{mac}$ to be large enough to effectively 
use the Law of Large Numbers (LLN) again \cite{macro}. Those fractions may also be thought of as 
representing macroscopic portions of the environment
accessible to multiple observers and  can be viewed as reflecting some sort of detection thresholds, e.g. the equipment sensitivity \cite{prl,pra}. 
Macrofraction states are given by the products:
\be
\varrho_t^{mac}(X)\equiv\bigotimes_{i\in mac}\varrho_t^{(i)}(X)
\ee
taken over all degrees of freedom in a given macrofraction. Since, crucially,  fidelity separates w.r.t. the tensor product, 
$B(\bigotimes_i\varrho_i,\bigotimes_i\sigma_i)=\prod_i B(\varrho_i,\sigma_i)$, one  easily obtains:
\be \label{Bmac}
B^{mac}_t(\Delta X)\equiv B[\varrho_t^{mac}(X),\varrho_t^{mac}(X')] = \prod_{i\in mac} B^i_t(\Delta X).
\ee
Crucially, the product here is taken over different degrees of freedom than those entering the decoherence factor (\ref{G}).
Here this is a fraction of the observed degrees of freedom, while in  (\ref{G}) those were the unobserved ones. 
This implies that the behavior of distinguishability $B^{mac}_t(\Delta X)$ and the partial decoherence factor  $\Gamma_t(\Delta X)$
is in general not correlated (as it would be if both quantities were calculated for the same degrees of freedom when $|\Gamma^{mac}_t(\Delta X)|\leq B^{mac}_t(\Delta X)$),
since each degree of freedom can have a different dynamics (here - a different frequency $\omega_i$) and couple differently to the central system (CM here). Thus,
both functions (\ref{G}) and (\ref{Bmac}) must be calculated independently \cite{prl}. In the short-time regime, we immediately obtain
from (\ref{b_st}):
\be
B^{mac}_t(\Delta X)\approx \text{exp}\left[-\frac{t^2}{8}\left(\frac{g\Delta X}{\hbar c^2}\right)^2 \sum_{i\in mac}F(\varrho_{0i};\hat H_{0i})\right]
\ee
Just like before, to make the expression more compact, let us introduce a probability $p_{mac}(\omega)$ such that the average QFI 
$\langle \langle F_{0}\rangle\rangle\equiv \int d\omega p_{mac}(\omega) F[\varrho_0(\omega);\hat H_{0\omega}] $ exists.
We note that now this probability is concentrated over the frequencies corresponding to a given macrofraction, which are in general
different from the ones that define the decoherence factor.
We obtain a formula similar to (\ref{Gst}) but with the average variance substituted for the average quantum Fisher information:
\be\label{Fst}
B^{mac}_t(\Delta X)\approx\text{exp}\left[-\frac{N_{mac} g^2\Delta X^2 \langle\langle F_{0} \rangle\rangle}{8 \hbar^2 c^4}t^2\right].
\ee
The exponential decay of this approximation with $N_{mac}$ can be viewed as a simple justification for the introduced coarse-graining:
While it may happen that the SBS (\ref{SBS0})-(\ref{perp}) is not formed on a "microscopic" level of single internal degrees of freedom,
it may still be approached at the coarse-grained level of  macrofractions.  
Expression (\ref{Fst}) defines the distinguishability time-scale in analogy to the decoherence time-scale (\ref{tdec}):
\begin{eqnarray}
\tau_{dst}\sqrt{N_{mac}\langle\langle F_0\rangle\rangle}\equiv \frac{\sqrt 8\hbar c^2}{g|\Delta X|}.\label{tdis}
\end{eqnarray}
Condition (\ref{st}) is fulfilled when the cumulative QFI is large compared to the relevant energies:
\be
N_{mac}\langle\langle F_0 \rangle\rangle\gg \left(8\hbar\omega\right)^2.\label{inf}
\ee
We note that since $F(\varrho_{0i};\hat H_{0i})\le 4\Delta H_{0i}^2$ \cite{geza}, $\tau_{dst}\ge\tau_{dec}$
if we consider  $N_\perp=N_{mac}$. Thus, it is faster to decohere than to concentrate information in the environment (cf. \cite{pomiary}).
A brief comment is in order: The formulas (\ref{Gst}, \ref{Fst}) are valid for arbitrary initial states of the internal degrees of freedom.
In the case of pure initial states, the state fidelity reduces to the state overlap so both expressions become identical functionally, although
they depend on different degrees of freedom. Thus, pure initial states, although fully captured by our analysis, are less interesting  than mixed ones.

\section{Spectrum Broadcast Structure formation for short times}
We now study the consequences of the above results for the extended state (\ref{mama}). Our main question is if the state  (\ref{mama}) is close to the SBS form (\ref{SBS0}).
We first note that the decoherence and distingushability times strongly depend on the height separation $|\Delta X|$. 
One can view (\ref{Gst}, \ref{Fst}) from another perspective: For a fixed time $t$ satisfying (\ref{st}), there are characteristic coherence and indistinguishability lengths:
\begin{eqnarray}
&&\Delta X_c^2\equiv \frac{2\hbar^2c^4}{g^2t^2 N_\perp \langle\langle\Delta H_{0}^2\rangle\rangle}\label{Dcoh},\\
&&\Delta X_d^2 \equiv \frac{8 \hbar^2c^4}{g^2t^2 N_{mac} \langle\langle F_0\rangle\rangle}\label{Ddst}
\end{eqnarray}
such that (cf. (\ref{Gst}, \ref{Fst})):

\begin{eqnarray}
&&\Gamma_t(\Delta X)\approx 0 \textrm{ for } |\Delta X| > \Delta X_c,\\
&&B_t(\Delta X)\approx 0 \textrm{ for } |\Delta X| > \Delta X_d.\label{dist}
\end{eqnarray}
We want to use this and approximate the double integral in (\ref{mama}) by discrete sums. We fix $t$, and change the 
integration variables in (\ref{mama}) to $\bar X\equiv (X+X')/2$ and $\Delta X=X'-X$ so that $X=\bar X-\Delta X/2$, $X'=\bar X+\Delta X/2$, and:
\be
\int\int dX dX' \equiv \int\int d\bar X d\Delta X.
\ee
Then, since $\Gamma_t(\Delta X)\approx 0$ for $|\Delta X |>\Delta X_c$, we can limit the $\Delta X$ integration range and obtain:
\begin{eqnarray}
\label{mama2}
&&\varrho_{ext}(t)=\int d\bar X \int_{-\Delta X_c}^{\Delta X_c} d\Delta X\nonumber\\
&&  \varrho_0\left(\bar X-\Delta X/2,\bar X +\Delta X/2\right)
\Gamma_t(\Delta X)\ket {\bar X-\Delta X/2}\bra {\bar X+\Delta X/2}\otimes\nonumber\\
&&\bigotimes_{i\in N_{o}} \hat U^{(i)}_t(\bar X-\Delta X/2)\varrho_{0i}\hat U^{(i)}_t(\bar X+\Delta X/2)^\dagger.
\end{eqnarray}
We now Taylor-expand the integrand in $\Delta X$, treating it as a small quantity. 
This can be done for the functions $\varrho_0\left(\bar X-\Delta X/2,\bar X +\Delta X/2\right)$, $\Gamma_t(\Delta X)$ and the operator $\hat U^{(i)}_t(\bar X-\Delta X/2)$
as they are all analytic functions of their arguments, but not with the vectors $\ket {\bar X-\Delta X/2}$. Thus, in the leading order the integral (\ref{mama2}) reads:
\begin{eqnarray}
&&\varrho_{ext}(t)=\nonumber\\
&& \int d\bar X  \varrho_0 (\bar X,\bar X )\int_{-\Delta X_c}^{\Delta X_c} d\Delta X \ket {\bar X-\Delta X/2}\bra {\bar X+\Delta X/2}\otimes\nonumber\\
&&\bigotimes_{i\in N_{o}} \hat U^{(i)}_t(\bar X)\varrho_{0i}\hat U^{(i)}_t(\bar X)^\dagger+O(\Delta X_c^2)\label{mama3}
\end{eqnarray}
(we note that the contribution from the  integration range $\bar X \sim \Delta X$ is of the order of $\Delta X_c^2$ and is thus included in the last term).
We now make use of the condition (\ref{dist}). We break the first integral into intervals of equal lengths $\Delta X_d$, centered at some  points $\bar X_k$: 
\be
\int d\bar X f(\bar X) =\sum_k \frac{1}{2}\int_{-\Delta X_d}^{\Delta X_d}  d\Delta \bar X f(\bar X_k+\Delta \bar X/2),
\ee
where $\Delta \bar X$ is the integration variable inside each interval. Assuming  $\Delta X_d$ to be small,
we may repeat the above approximation, expanding everything that is smooth in $\bar \Delta X$ and keeping the lowest term. We obtain: 
\begin{eqnarray}
&&\varrho_{ext}(t)=\nonumber\\ 
&&\frac{1}{2}\sum_k p_0(\bar X_k)\int_{-\Delta X_d}^{\Delta X_d}   d\Delta \bar X\int_{-\Delta X_c}^{\Delta X_c}d\Delta X
\left|\bar X_k+\frac{\Delta \bar X-\Delta X}{2}\right\rangle\left\langle \bar X_k +\frac{\Delta \bar X+\Delta X}{2}\right| \otimes
\nonumber\\
&&\bigotimes_{i\in N_{o}}\varrho_t^{(i)}(\bar X_k) +O(\Delta X_c^2,\Delta X_d^2 ),\label{mama4}
\end{eqnarray}
where $p_0(X)\equiv\varrho_0(X,X)=\bra X \varrho_0\ket X$ and we used (\ref{rhot}). 
Changing once again the integration variables to sum and difference of $\Delta X, \Delta \bar X$, $\Delta_\pm\equiv (\Delta X \pm\Delta \bar X)/2$: 
\be
\frac{1}{2}\int d\Delta \bar X\int d\Delta X \equiv \int d\Delta_-\int d\Delta_+
\ee
we obtain:
\begin{eqnarray}
&&\varrho_{ext}(t)=\nonumber\\ 
&&\sum_k p_0(\bar X_k) \int_{-\bar\Delta}^{\bar\Delta} d \Delta_+ \int_{-\bar\Delta}^{\bar\Delta}d \Delta_-
\ket {\bar X_k+\Delta_-}\bra {\bar X_k + \Delta_+}\otimes
\nonumber\\
&&\bigotimes_{i\in N_{o}}\varrho_t^{(i)}(\bar X_k) +O(\Delta X_c^2,\Delta X_d^2 ),\label{mama5}
\end{eqnarray}
where $\bar\Delta\equiv (\Delta X_c+\Delta X_d)/2$. Note that the integrals in $\Delta_\pm$ are separated now which allows us to introduce the smeared position states: 
\be
\ket{\bar X;\bar\Delta }\equiv\int_{-\bar\Delta}^{\bar\Delta}d\Delta X \ket {X+\Delta X}.
\ee
Using them and the partition of $N_o$ into the macrofractions, we can finally rewrite (\ref{mama5}) as: 
\begin{eqnarray}
&&\varrho_{ext}(t)=\label{SBS}\\
&&\sum_k p_0(\bar X_k)\ket{\bar X_k;\bar\Delta }\bra{\bar X_k;\bar\Delta}\otimes\varrho_t^{mac_1}(\bar X_k)\otimes\dots\otimes\varrho_t^{mac_{\mathcal M}}(\bar X_k)+O(\Delta X_c^2,\Delta X_d^2)
\nonumber
\end{eqnarray}
The macrofraction states $\varrho_t^{mac}(\bar X_k)$ are 
by the above construction almost perfectly distinguishable for different $\bar X_k$, since their separation is  $\sim\Delta X_d$. 
Thus, the CM position is decohered to within $\bar \Delta$ 
and is  stored in the internal degrees of freedom in many identical copies. The structure (\ref{SBS}) is a version of the
Spectrum Broadcast Structure for continuous variables. Please note its coarse-grained character in the variable $X$.
Our analysis here is somewhat heuristic.
Mathematically rigorous proof would consist in showing how fast the actual extended state approaches some ideal SBS
depending on the decoherence and the fidelities.  We will postpone this non-trivial mathematical question to a more specialized publication,
pointing that for finite dimensional systems this question was solved in \cite{piotrek}. It is shown there that indeed 
the trace distance of the actual extended state to some ideal SBS is bounded by the decoherence factor and
the macroscopic fidelities.

\section{An example of information encoding: Displaced thermal environments}
As an illustration of the above general reasoning, we consider the the internal degrees of freedom to be prepared in thermal coherent states:
\be
\varrho_{0i}=\hat D(\alpha)\varrho^{(i)}_{th}\hat D(\alpha)^\dagger,\label{init}
\ee
where $\hat D(\alpha)$ is the displacement operator (assumed here the same for all the oscillators)
and $\varrho^{(i)}_{th}\equiv \textrm e^{-\beta\hat H_{0i}}/Z_i$, $Z_i\equiv \tr(\text e^{-\beta\hat H_{0i}})$, $\beta\equiv 1/(kT)$.
Those states no longer commute with $\hat H_{0i}$ and we have:
\begin{eqnarray}
&&\Delta  H_{0i}^2=(\Delta  H_{0i}^2)_{th}+(\hbar\omega_i|\alpha|)^2\mathrm{cth}\left(\frac{\beta\hbar\omega_i}{2}\right),\\
&& F(\varrho_{0i};\hat H_{0i})= 4 (\hbar\omega_i|\alpha|)^2 \mathrm{th}\left(\frac{\beta\hbar\omega_i}{2}\right),
\end{eqnarray} 
where $\langle \hat A\rangle_{th}\equiv \tr(\varrho_{th}\hat A)$ and  $\mathrm{cth(\cdot)}, \mathrm{th(\cdot)}$ stand for the hyperbolic cotangent and the hyperbolic tangent respectively. 
As a digression, we note that the reciprocal dependence on the temperature of the non-thermal parts above,  
leads to a formal relation:
\be
[\Delta  H_{0i}^2-(\Delta  H_{0i}^2)_{th}]F(\varrho_{0i};\hat H_{0i})=4(\hbar\omega_i|\alpha|)^4.\label{relation}
\ee
In light of the short-time expressions (\ref{G_st}) and (\ref{b_st}), it may be interpreted as a type of an information gain-vs.-disturbance relation (cf. e.g. \cite{michal}).
Here, the disturbance to the central system (in a form of a non-thermal contribution to the decoherence) 
is characterized by $\Delta  H_{0i}^2-(\Delta  H_{0i}^2)_{th}$, while the environment information gain 
by $F(\varrho_{0i};\hat H_{0i})$. The meaning of (\ref{relation})  here is that the hotter the environment is, the stronger is its decohering power
but the lesser is its information capacity \cite{epl}. One should stress  that (\ref{relation}) applies only for a specific case of 
displaced thermal states. Its validity and form for more general states and relation to known results of a similar kind \cite{natcom} will be studied elsewhere.

\begin{figure}[t]
	\centering
	\subfigure[ single oscillator]{
		\includegraphics[width=0.5\textwidth]{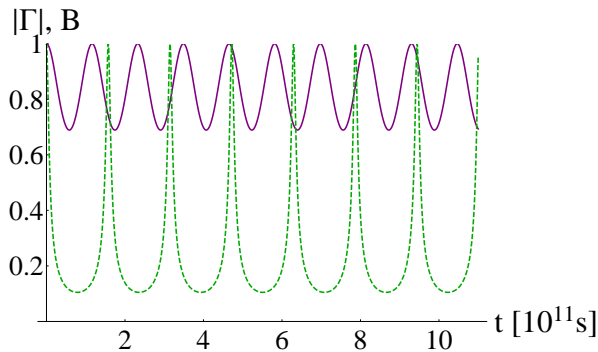}
	}

	\subfigure[1000 oscillators]{
		\includegraphics[width=0.5\textwidth]{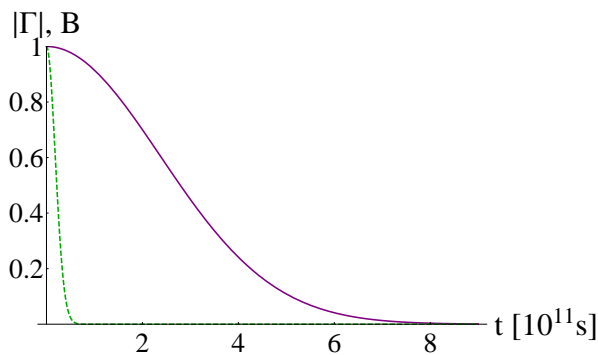}
	}
	\caption{(Color online) Time dependencies of $|\Gamma_t(\Delta X)|$ (green dotted line)  and $B_t(\Delta X)$  (solid magenta) for intermediate temperatures $T=10$K ($k_BT\approx 3\hbar\omega_{max}$).
		Plot (a) shows a single-oscillator case, where the observed and the unobserved fractions consist of a single oscillator each.  The unobserved (traced out) oscillator has a frequency of $\omega_\perp=3.6\times10^{11}$s$^{-1}$,
             while the observed one $\omega_o=4.9\times10^{11}$s$^{-1}$, which clearly demonstrate that they couple differently to the center of mass and it is possible that  $|\Gamma_t(\Delta X)|\approx 1$, but $B_t(\Delta X)<1$. 
             Plot (b) shows $N_\perp=N_{mac}=10^3$ random oscillators, 
		drawn independently for the observed and the unobserved fractions from an uniform distribution over $[1\dots 5]\times 10^{11}$s$^{-1}$.  
		The other parameters are $\Delta X=10^{-6}$m, $|\alpha|=1$, common to both plots.
		Note  the extremely long decay times in (b) due to the small macrofraction sizes as compared to the gravitational interaction strength. }
	\label{time_dep}
\end{figure}

One can go beyond the above short-time analysis and give  
a compact and exact expressions for $|\Gamma_t(\Delta X)|$ and $B_t(\Delta X)$ for arbitrary times. 
A direct calculation  for a single band $\omega_i$ leads to:
\begin{eqnarray}\label{Gt}
&&|\Gamma_t^i(\Delta X)|=|\Gamma^{th}_t(\Delta X)|\text{exp}\left[-|\alpha|^2|\Gamma^{th}_t(\Delta X)|^2\mathrm{cth}\left(\frac{\beta\hbar\omega_i}{2}\right)\left(1-\cos\Delta\phi_i\right)\right],\nonumber\\
&&
\end{eqnarray}
where $|\Gamma^{th}_t(\Delta X)|\equiv [1+2\bar n_i(\bar n_i+1)(1-\cos\Delta\phi_i)]^{-1/2}$, 
$\bar n_i\equiv \langle\hat n_i\rangle_{th}$. Calculation of $B^i_t(\Delta X)$ is more involved, we perform them in \ref{app:fid}, where we find:
\be
B_t^i(\Delta X)=\text{exp}\left[-|\alpha|^2\mathrm{th}\left(\frac{\beta\hbar\omega_i}{2}\right)\left(1-\cos\Delta\phi_i\right)\right]. \label{Bt}
\ee
The above single-band functions are periodic in time (hidden in $\Delta\phi_i$ (\ref{st})) and obviously there is no SBS formation at the level
of single environments. However, for macrofractions, described by a collection of randomly distributed $\omega_i$ \cite{epl}, the corresponding decoherence and fidelity
factors become quasi-periodic functions of time:
\begin{eqnarray}
&&|\Gamma_t(\Delta X)|=\nonumber\\
&&\prod_{i=1}^{N_\perp}|\Gamma^{th}_t(\Delta X)|\text{exp}\left[-|\alpha|^2|\Gamma^{th}_t(\Delta X)|^2\mathrm{cth}\left(\frac{\beta\hbar\omega_i}{2}\right)\left(1-\cos\Delta\phi_i\right)\right],\\
&&B_t^{mac}(\Delta X)=\text{exp}\left[-|\alpha|^2\sum_{i=1}^{N_{mac}}\mathrm{th}\left(\frac{\beta\hbar\omega_i}{2}\right)\left(1-\cos\Delta\phi_i\right)\right],
\end{eqnarray}
and the random phases $\Delta\phi_i$ may lead to their effective damping. This, however, depends on the temperature (apart from the other factors kept fixed)
as e.g. for a high temperature $\frac{k_BT}{\hbar \omega_i}\to \infty$, $|\Gamma_t^i(\Delta X)|=O\left(\frac{\hbar \omega_i}{k_BT}\right)$ while $ B_t^i(\Delta X)=1-O\left(\frac{\hbar \omega_i}{k_BT}\right)$.
For a low temperature in turn, the initial states (\ref{init}) become pure and $B_t^i(\Delta X)=|\Gamma_t^i(\Delta X)|$,
which follows from  $B(\ket\psi\bra\psi, \ket\phi\bra\phi)=|\langle\psi|\phi\rangle|$. 
An example of the intermediate regime $k_BT\approx\hbar \omega_i$ is shown in Fig.~\ref{time_dep} . We see that for big enough
macrofraction sizes (cf. \cite{jp}), both decoherence factor and fidelity decay, indicating the SBS formation. We note a very long, compared to usual decoherence times,
time-scale of this process, caused by the weak nature of the gravitational coupling. The times become much shorter for macroscopic ($N\sim 10^{23}$),
rather than mesoscopic fraction sizes. 

\section{Conclusions}
We have analyzed information transfer to the internal degrees of freedom during the universal gravitational decoherence of \cite{pikovski},
using the recently developed methods of Spectrum Broadcast Structures \cite{prl,pra}. We have shown that generically, on short time scales
there is an SBS formation, implying a redundant encoding of the center-of-mass position in the internal degrees of freedom during the decoherence. 
Importantly, we have derived the relevant time-scales of  the process, given by the energy dispersion and the Quantum Fisher Information. The role of the latter in the SBS formation 
have not been recognized before. The resulting SBS has a natural coarse-grained structure, appropriate for the continuous variable case,
reflecting finite coherence and distinguishability lengths. 

Going beyond the short-time analysis, we have analyzed the case when the internal degrees of freedom are prepared in a displaced thermal states.
This breaks the symmetry with respect to the free internal evolution and allows for the information encoding in the environment, unlike
in the purely thermal case of \cite{pikovski}. We have shown that for big enough macrofraction sizes, there are temperature regimes where the 
SBS is being formed. As a by-product of this analysis we have derived a form of an information gain-vs.-disturbance relation (\ref{relation}).

We finish with some speculations on the analogy between the above information transfer process and the, so called, point individuation in General Relativity.
As emphasized by the famous Einstein's Hole argument, active diffeomorphism invariance 
of General Relativity forbids assigning a physical meaning not only to coordinate charts but to very (mathematical) manifold points as well.
This has led to a still active debate over the ontology of space-time and physical objectivity of space-time points \cite{stachel1, meets}.
A way out of the situation is to define space-time points using coincidences of matter, e.g. as intersection points of particles' world-lines.
This process is called space-time point individuation (or objectivization), and
the degrees of freedom used are called individuating fields.
The peculiarity of the theory is that individuation can be in principle achieved in a fully dynamical way 
by the metric field itself, provided it satisfies the Einstein equations \cite{stachel2}.

On the other hand, the SBS structure (\ref{SBS}) encodes a certain form of point objectivity as well: 
Indirect observation of the position through 
any of the environmental macrofractions
will always give the same result $\bar X_k$, leaving the CM located around $\bar X_k$ and without any disturbance, 
on average, to the global state.  This observer-invariance and non-disturbance (which may be viewed as 
a type of a time-invariance) can be taken as a basis of an operational definition 
of objectivity \cite{ZurekNature}, in this case of the (approximate) position. 
Moreover, it can be shown that under some general assumptions the only state structure compatible with so defined objectivity
is precisely the SBS \cite{pra} and that it is generic for macroscopic quantum measurements \cite{pomiary}. 
An intriguing question thus arises if the gravitational decoherence mechanism 
may also lead, through the SBS, the the point individuation? This would be an example of a quantum individuation.

There are of course some obvious differences. First of all, described here gravity-generated
SBS provides objectivization of points of space only, not of space-time events (see however \cite{penrose}). But we have used
just a first correction to the non-relativistic  Schr\"odinger equation \cite{lamerzahl}  rather than e.g. Tomonaga-Schwinger evolution law,
which could be a future direction. Second, by the very construction there is a finite precision (\ref{Dcoh}) and resolution (\ref{Ddst}) 
with which a point can be localized. Both of these parameters parameters improve with the macrofraction size, which suggests
that objective points may be macroscopic phenomena and the objectivity breaks down at some microscopic scales
(cf. macro-objectivity idea \cite{photonics}). Answers to these questions may shed some light on what happens with 
the space-time at microscopic scales.
\section{Acknowledgments}

We would like to thank  K. Rza\.{z}ewski, R. Horodecki and P. Horodecki for discussions. We acknowledge the financial support
of the John Templeton Foundation through the grant ID \#56033. 

\appendix

\section{Fidelity for displaced thermal states}
\label{app:fid}
Here we calculate $B[\varrho_t^{(i)}(X),\varrho_t^{(i)}(X')]$, where
\ben
\varrho_t^{(i)}(X)= U^{(i)}_t(X)\varrho_{0i}U^{(i)}_t(X')^\dagger,
\label{eqap:1}
\een
$\varrho_{0i}=\hat D(\alpha)\varrho^{(i)}_{th}\hat D(\alpha)^\dagger$ is the displaced thermal state and  $U^{(i)}_t(X)=\text{exp}[-i\omega_i(X)t\hat n_i]$. We start by rewriting the state as
\ben
\label{eqap:2}
&&\varrho_t^{(i)}(X)= \hat D(\alpha e^{-i\omega_i(X)t} )\varrho^{(i)}_{th}\hat D(\alpha e^{-i\omega_i(X)t})^\dagger.
\een
To arrive at the above expression we inserted in  (\ref{eqap:1}) the resolution of identity $I= U_t^{(i)}(X)^\dagger U_t^{(i)}(X)$ between the left displacement operator and the state  and its conjugate version between the state and right, conjugated, displacement operator. Then we used the fact that: i) the unitary commutes with the thermal state ii) $ U_t^{(i)}(X) \hat D(\alpha ) U_t^{(i)}(X)^\dagger = \hat D(\alpha e^{-i\omega_i(X)t} )$. Dropping dependence on $i$ and using (\ref{eqap:1}) we obtain:
\begin{equation}\label{Bmic}
B[\varrho_t(X),\varrho_t(X')]=tr\sqrt{\sqrt{\varrho_{th}} \hat D(\eta_t) \varrho_{th} \hat D(\eta_t)^\dagger \sqrt{\varrho_{th}}},\\
\end{equation}
where 
\begin{equation}
\hat D(\eta_t) \equiv \hat D(\alpha (e^{-i\omega_i(X')t}-e^{i\omega_i(X)t})) ,\label{et}
\end{equation}
and we have pulled the extreme left and right displacement operators out of both the square roots and used the cyclic property of the trace to cancel them out.
The phase factors resulting from composition of remaining displacement operators cancel out as both unitary 
operators under the square root are Hermitian conjugates of each other.
Next,  we use the corresponding $P$-representation for the middle $\varrho_{th}$ under the square root in (\ref{Bmic}):
\begin{equation}\label{ThermalInitialCnditions}
\varrho_{th}(\bar{n})\equiv\frac{1}{\bar n}\int\frac{d^2\gamma}{\pi}e^{-\frac{|\gamma|^2}{\bar n}}\ket\gamma\bra\gamma, 
\end{equation}
where $\bar n=1/(e^{\beta \omega}-1)$, $\beta\equiv\hbar/(k_B T)$. 
Indicating the Hermitian operator under the square root in (\ref{Bmic}) by $\tilde B_t$, we obtain:
\begin{eqnarray}\label{B1}
\tilde{B}_t=&\int\frac{d^2\gamma}{\pi\bar n}e^{-\frac{|\gamma|^2}{\bar n}}\sqrt{\varrho_{th}} \hat D(\eta_t)\ket\gamma\bra\gamma \hat D(\eta_t)^\dagger\sqrt{\varrho_{th}}\nonumber\\
=&\int\frac{d^2\gamma}{\pi\bar n}e^{-\frac{|\gamma|^2}{\bar n}}\sqrt{\varrho_{th}}\ket{\gamma+\eta_t}\bra{\gamma+\eta_t}\sqrt{\varrho_{th}}.
\end{eqnarray}
The next step is to calculate  explicitly the square root in the equation above. For this aim we expand $\rho_0$ in the Fock basis:
\begin{equation}\label{FockRepresentation}
\varrho_{th}=\sum_k\frac{\bar n^k}{(\bar k +1)^{n+1}}\ket k\bra k.
\end{equation}
Replacing it in Eq.\ (\ref{B1}) we have:
\begin{equation}\label{B2}
\tilde{B}_t=\int\frac{d^2\gamma}{\pi\bar n}e^{-\frac{|\gamma|^2}{\bar n}}\sum_{i,j}\lambda_{ij}(\bar n)
{\langle j|\gamma+\eta_t\rangle}{\langle \gamma+\eta_t|i\rangle}\ket{j}\bra{i}
\end{equation}
with:
\begin{equation}
\lambda_{ij}(\bar n)\equiv\sqrt{\frac{\bar n^{i+j}}{(\bar n +1)^{i+j+2}}}.
\end{equation}
Using the Fock basis $\ket{j}$ representation of coherent states one may explicit the scalar product $\langle j|\gamma+\eta_t\rangle$. Accordingly Eq.\ (\ref{B2}) gets:
\begin{eqnarray}
\tilde{B}_t&=\frac{1}{\bar n +1}e^{-\frac{|\eta_t|^2}{1+2\bar n}}\int\frac{d^2\gamma}{\pi\bar n}e^{-\frac{1+2\bar n}{\bar n(\bar n +1)}
	\left|\gamma+\frac{\bar n}{1+2\bar n}\eta_t\right|^2}\times\nonumber\\
&\times \left|\sqrt{\frac{\bar n}{\bar n +1}}(\gamma+\eta_t)\right\rangle\left\langle\sqrt{\frac{\bar n}{\bar n +1}}(\gamma+\eta_t)\right|.\label{Btilde}
\end{eqnarray}
We now show that this equation is formally equivalent to that of a thermal state introduced in Eq.\ (\ref{ThermalInitialCnditions}). 
For this aim, we underline that we are interested in the square root of the operator  $\tilde{B}_t$, rather than in itself. Therefore, there is a freedom of rotating $\tilde{B}_t$ by a unitary operator, and in particular a displacement one:
\begin{equation}\label{DroppingDisplacement}
\text{Tr}\left[\sqrt{ \hat D\tilde{B}_t \hat D^\dagger}\right]=
\text{Tr}\left[\hat D\sqrt{\tilde{B}_t} \hat D^\dagger\right]=
\text{Tr}\left[\sqrt{\tilde{B}_t}\right].
\end{equation}
In particular we find:
\begin{eqnarray}\label{DisplacementShift}
&  \left|\sqrt{\frac{\bar n}{\bar n +1}}(\gamma+\eta_t)\right\rangle\varpropto \hat D\left(\sqrt{\frac{\bar n}{\bar n +1}}\frac{1+\bar n}{1+2\bar n}\right) \left| \sqrt{\frac{\bar n}{\bar n +1}}(\gamma+\frac{\bar{n}}{1+2\bar{n}}\eta_t) \right\rangle,
\end{eqnarray}
where we have omitted the irrelevant phase factor arising from the action of the displacement. 
We replace Eq.\ (\ref{DisplacementShift}) into Eq.\ (\ref{Btilde}). Dropping displacement operators due to Eq.\ ($\ref{DroppingDisplacement}$) and introducing the variable:
\begin{equation}
\tilde{\gamma}=\sqrt{\frac{\bar n}{\bar n +1}}\left(\gamma+\frac{\bar n}{1+2\bar n}\eta_t\right)
\end{equation}
one obtains:
\begin{equation}
B[\varrho_t(X),\varrho_t(X')]=\frac{e^{-\frac{1}{2}\frac{|\eta_t|^2}{1+2\bar n}}}{\sqrt{1+2\bar n}}\text{Tr}\sqrt{\rho_{th}\left(\frac{\bar n^2}{1+2\bar n}\right)}.
\end{equation}
In order to calculate  explicitly  the square root we recall the Fock expansion in Eq.\ (\ref{FockRepresentation}).
Finally, we get:
\begin{equation}
B[\varrho_t^{(i)}(X),\varrho_t^{(i)}(X')]=\text{exp}\left[-|\alpha|^2\mathrm{th}\left(\frac{\beta\hbar\omega_i}{2}\right)\left(1-\cos\Delta\phi_i\right)\right]. \label{Bt}
\end{equation}


\begin{thebibliography}{99}

\bibitem{decoh}    Joos E, Zeh H D, Kiefer C, Giulini D J W, Kupsch J, Stamatescu I-O 2003
\emph{Decoherence and the Appearance of a Classical World in Quantum Theory} (Berlin: Springer),  Schlosshauer M 2007 \emph{Decoherence and the Quantum-to-Classical Transition}
(Berlin: Springer).

\bibitem{pikovski} Pikovski I,  Zych M,  Costa F, and  Brukner \v C 2015 Universal decoherence due to gravitational time dilation  \emph{Nature Phys.} {\bf 11}, 668.

\bibitem{natcom}  Zych M, Costa F, Pikovski I, and Brukner 2011 Quantum interferometric visibility as a witness of general relativistic proper time \emph{Nat. Commun.} \textbf{2}, 505.

\bibitem{gravitation}  Misner C W, Thorne K S, and Wheeler J A, 1973 \emph{Gravitation},  	(W. H. Freeman: San Francisco).

\bibitem{diosi}  Di\'osi L 1987 A universal master equation for the gravitational violation of quantum mechanics  \emph{Phys. Lett. A} {\bf 120}, 377.

\bibitem{penrose} Penrose R 1996 On Gravity's role in Quantum State Reduction \emph{Gen. Relativ. Gravit.} {\bf 28}, 581.

\bibitem{bonder1} Bonder Y, Okon E, and Sudarsky D 2016 Questioning universal decoherence due to gravitational time dilation \emph{Nature Phys.} {\bf 12}, 2.

\bibitem{bonder2}  Bonder Y, Okon E, and Sudarsky D 2015 Can gravity account for the emergence of classicality? \emph{Phys. Rev. D} {\bf 92}, 124050.

\bibitem{diosi2} Di\'osi L 2015 Centre of mass decoherence due to time dilation: paradoxical frame-dependence   arXiv:1507.05828.

\bibitem{qa}  Pikovski I,  Zych M,  Costa F, and  Brukner \v C 2015 Time Dilation in Quantum Systems and Decoherence: Questions and Answers arXiv:1508.03296.

\bibitem{pang}  Pang B H,  Chen Y, and Khalili F YA 2016 Universal Decoherence under Gravity: A Perspective through the Equivalence Principle \emph{Phys. Rev. Lett.} {\bf 117}, 090401. 

\bibitem{pr}  Pound R V and   Rebka G A Jr 1960 Apparent Weight of Photons \emph{Phys. Rev. Lett.} {\bf 4}, 337.

\bibitem{wineland} Chou C W, Hume D B, Rosenband T, and Wineland D J 2010 Optical Clocks and Relativity \emph{Science} {\bf 329}, 1630.

\bibitem{adler} Adler S L and Bassi A 2016 Gravitational Decoherence for Mesoscopic Systems \emph{Phys. Lett. A} {\bf 380}, 390. 


\bibitem{ZurekNature}  Zurek W H 2009 Quantum Darwinism \emph{Nature} {\bf 5} 181.

\bibitem{prl} Korbicz J K, Horodecki P and Horodecki R 2014 Objectivity in a Noisy Photonic Environment through
Quantum State Information Broadcasting  \emph{Phys. Rev. Lett.}  {\bf 112} 120402.

\bibitem{pra}  Horodecki R, Korbicz J K and Horodecki P 2015 Quantum origins of objectivity \emph{Phys. Rev. A}  \textbf{91} 032122. 

\bibitem{lamerzahl}  L\"ammerzahl, C 1996 A Hamilton operator for quantum optics in gravitational fields \emph{ Phys. Lett. A} 203, 12.

\bibitem{pawel} 
Brandao F G S L,  Piani M Horodecki P 2015 Generic emergence of classical features in quantum Darwinism  \emph{Nat. Comm.} \textbf{6} 7908F.

\bibitem{geza} Tóth G and  Apellaniz I 2014 Quantum metrology from a quantum information science perspective  \emph{J. Phys. A: Math. Theor.} {\bf 47}, 424006.

\bibitem{stachel1} Stachel J 2014 \emph{Living Rev. Relativity} 2014 The Hole Argument and Some Physical and Philosophical Implications {\bf 17}, 1.


\bibitem{Zwolak}  Zwolak M,  Riedel C J, and  Zurek W H 2014 Amplification, Redundancy, and Quantum Chernoff Information Phys. Rev. Lett. {\bf 112}, 140406.

\bibitem{pomiary}  Korbicz J K,  Aguilar E A,  \'Cwikli\'nski P and  Horodecki P  2017 Do objective results typically appear in quantum measurements? Phys. Rev. A {\bf 96}, 032124.

\bibitem{fuchs} Fuchs C A and van de Graaf J 1999 Cryptographic Distinguishability Measures for Quantum-Mechanical States   \emph{IEEE Trans. on Inf. Theor} {\bf 45} 1216. 

\bibitem{Chernoff} Calsamiglia J, and Mu\~noz-Tapia R, Masanes Ll, Acin A, and Bagan E 2008 Quantum Chernoff bound as a measure of distinguishability between density matrices: Application to qubit and Gaussian states \emph{Phys. Rev. A} {\bf 77}, 032311 

\bibitem{Zyczkowski_ksiazka} Bengtsson I and \.Zyczkowski K 2016 \emph{Geometry of Quantum States: An Introduction to Quantum Entanglement} (Cambridge: Cambridge University Press).

\bibitem{qfi}  Braunstein S L and Caves C M Statistical distance and the geometry of quantum states 1994  \emph{Phys. Rev. Lett.} {\bf 72}, 3439.

\bibitem{metrology} Giovannetti V, Lloyd S, and Maccone L  Advances in quantum metrology 2011 \emph{Nature Photon.} {\bf 5}, 222.

\bibitem{macro} This coarse graining, introduced in the present context in \cite{prl}, is analogous to e.g. description of liquids:
Each point of a liquid (a macro-fraction here) is in reality composed of a suitable large number of 
microparticles (individual oscillators). Same technique is also used in a mathematical approach to von Neumann
measurements using, so called, macroscopic observables;
see e.g.  Sewell G 2015 On the mathematical structure of quantum measurement theory \emph{Rep. Math. Phys.} {\bf 56}, 271  and the references therein.

\bibitem{michal} Buscemi F, Hayashi M, and Horodecki M 2008 Global information balance in quantum measurements  \emph{Phys. Rev. Lett.} {\bf 100}, 210504.

\bibitem{piotrek}  Mironowicz P, Korbicz J K, and Horodecki P 2017 Monitoring of the process of system information broadcasting in time, \emph{Phys. Rev. Lett.} {\bf 118}, 150501.

\bibitem{epl} Tuziemski J and Korbicz J K 2015 Dynamical objectivity in quantum Brownian motion  \emph{EPL (Europhysics Letters)} {\bf 112} 40008.

\bibitem{jp} Zych M, Pikovski I, Costa F, and Brukner \v C 2016 General relativistic effects in quantum interference of "clocks" \emph{J. Phys.: Conference Series} {\bf 723}, 012044.

\bibitem{meets}  Callender C and  Huggett N eds 2001 \emph{Physics meets philosophy at the Planck scale: Contemporary theories in quantum gravity.},
(Cambridge: Cambridge University Press).

\bibitem{stachel2} Stachel J, in  Earman J,
 Janis I, Massey G J, and Rescher N eds 1993 \emph{Philosophical problems of the internal and
	external worlds, essays on the philosophy of Adolf Gr\"unbaum}, (Pittsburgh: University of Pittsburgh Press),
 Lusanna L and Pauri M 2006 
 Explaining Leibniz-equivalence away: Dis-solution of the Hole Argument and physical individuation of point  \emph{Stud. Hist. Phil. Mod. Phys.} {\bf 37}, 692.


\bibitem{photonics} Tuziemski J and Korbicz J K 2015 Objectivisation In Simplified Quantum Brownian Motion Models \emph{Photonics} \textbf{2(1)} 228  















\end{thebibliography}
\end{document}